\documentclass[onecolumn,11pt]{article}
\usepackage{amssymb,amsmath,epsfig}
\usepackage{epstopdf}
\begin{document}

\title{\bf Holographic application in cosmology: Thermodynamics of the Van der Waals cosmic fluid}
\author{\bf{Mahasweta Biswas$^1$}\thanks {mahasweta.bsc11@gmail.com}~,
\bf{Sayani Maity$^2$}\thanks{sayani.maity88@gmail.com} ~and
\bf{Ujjal Debnath$^1$}\thanks{ujjaldebnath@gmail.com}\\\\
$^1$Department of Mathematics, Indian Institute of Engineering\\
Science and Technology, Shibpur, Howrah-711 103, India.\\
$^2$Department of Mathematics, Techno India Salt Lake, Sector-V,\\
Kolkata-700 091, India.}

\date{\today}
\maketitle

\begin{abstract}
This paper is devoted to investigate the thermodynamic stability
of a generic cosmological fluid known as Van der Waals fluid in
the context of flat FRW universe. It is treated as a perfect fluid
that obeys the equation of state
$P=\frac{\gamma\rho}{1-\beta\rho}-\alpha\rho^{2}, 0\leq\gamma<1$,
where $\rho$ stands for energy density and $P$ stands for pressure
of the fluid. In this regard, we discuss the behavior of physical
parameters to analyze the evolution of the universe. We
investigate whether the cosmological scenario fulfills the third
law of thermodynamics using specific heat formalism. Next we
discuss the thermal equation of state and by means of adiabatic,
specific heat and isothermal conditions from classical
thermodynamics we examine the thermal stability.

\end{abstract}


\section{Introduction}
The observations like Type Ia supernova
\cite{1RAGFAVCA98,1PSAGGGKRA99,1BPAPARBJJBJR00,1PSAGARAPBG03}, the
large scale structure
\cite{1CMDGMSSWNPCS01,1TMSMABMRAKDSSHW04,1CSPWJPRANP05,1SVFCSWSMD06},
cosmic microwave background (CMB)
\cite{1HSAPBABJBA00,1NCBAPARBJJBJR02,1SDNVLPHVKE03} indicate that
the universe endures an accelerating expansion. The source of this
phenomenon is suspected to be an exotic fluid that violates strong
energy condition and possesses a large negative pressure, dubbed
as dark energy. According to Planck's observational data
\cite{1AUSVSAA04} about 68.3 percent of the total cosmic budget is
occupied by Dark energy, while about 26.8 percent is filled with
dark matter and about 4.9 percent is usual baryonic matter.There
have been a prolonged attempt to reconcile the physical nature of
dark energy. An overwhelming flood of dynamical dark energy models
such as quintessence, tachyon \cite{1SA02}, ghost
\cite{1MMNTPF15}, k-essence \cite{1CTOTYM00}, fermionic field
\cite{1MR10,1MSDU13,1MSDU14}, phantom \cite{1NSOSD03}, Chaplygin
gas \cite{1PVMUKAYMUPV01}, holographic dark energy (HDE)
\cite{1WS98,1CEJSMTS06,1PT03}, new agegraphic dark energy (NADE)
\cite{1WHCRG081,1MSDU16} and modified gravity models such as f(R)
gravity \cite{1CSCSTA,1CSMDVTMTMS04}, f(T) gravity
\cite{1FRFF07,1BGRFR09,1LEV10}, Ho$\check{r}$ava-Lifshitz gravity
\cite{Horava1, Horava2,
Horava3,MSRP181} have been proposed in literature.\\

Like Chaplygin gas family, another single-component fluid known as
Van der Waals fluid \cite{Cap} has attracted much attention for
unification of different fluid. The main feature of this model is
to reproduce the accelerated and matter-dominated phases with a
single component. The perfect fluid equation of state is not
realistic in describing the all the phases of the evolution of the
universe. The properties of Van der Waals fluid have been analyzed
in \cite{1JRCSCMHB16}. It is indeed the holographic description of
the dark energy, so by having information about the Van der Waals
fluid we can obtain knowledge about the accelerating expansion of
universe. In \cite{1KGM03,1KGM04} a mixture of two fluids, taking
the Van der Waals fluid as dark energy and the perfect gas
equation of state for the matter, have been considered to get a
whole dynamics of the universe. In Ref. \cite{1KMPBCE13}, a toy
model of the Universe is considered with generalized ghost dark
energy, Van der Waals fluid and some modified fluids. They have
studied the unusual connection among different fluids. Although
there have been a lot of work \cite{1CVFTCTACS06,1SB06,1KMPBKEO14}
discussing various aspects of Van der Waals fluid in the standard
cosmological model with observations, there has not been done a
full thermodynamic constraints of Van der Waals fluid.
Thermodynamics of Chaplygin gas model has been studied by Myung
\cite{Myung}. Santos et al \cite{Santos1,Santos2} have studied the
thermodynamic stability of the generalized and modified Chaplygin
gas models. Thermodynamics of Modified Chaplygin Gas and Tachyonic
Field have been analyzed in ref. \cite{Bhatta}. Thermodynamic
stability of generalized cosmic Chaplygin gas has been studied by
Sharif et al \cite{Shar}. Motivated by these works, here, we
examine the thermodynamic
stability of Van der Waals fluid in the background of a flat FRW universe.\\

The paper is organized as follows. In section 2, we study the
behavior of pressure, EoS parameter, deceleration parameter and
also analyze the stability using the sign of square speed of
sound. In section 3, we study the thermodynamic stability of Van
der Waals fluid. We devote the last section for summarization of
the results.

\section{Physical Features of Van der Waals Fluid}\label{sec2}

Here we assume the flat Friedmann-Robertson-Walker (FRW) model of
the universe represented by the following line element:
\begin{eqnarray}
ds^{2}=-dt^{2}+a^{2}(t)(dr^{2}+r^{2}d\theta^{2}+r^{2}sin^{2}\theta
d\phi^{2}),
\end{eqnarray}
where $a(t)$ is the scale factor. Now we assume the equation of
state for Van der Waals fluid as in the form \cite{Cap}
\begin{eqnarray}
P=\frac{\gamma\rho}{1-\beta\rho}-\alpha\rho^{2}, 0\leq\gamma<1,
\alpha=3p_{c}\rho_{c}^{-2},\beta=(3\rho_{c})^{-1},
\end{eqnarray}
where $\rho$ and $P$ represent the energy density and pressure of
the Van der Waals fluid. Here $\rho_{c}$ and $p_{c}$ represnet the
critical density and the critical pressure of the Van der Waals
fluid at critical point. The above equation reduces to the perfect
fluid case in the limit $\alpha,\beta\rightarrow 0$. The energy
density of the fluid can be written in the form:
\begin{eqnarray}
\rho=\frac{U}{V},
\end{eqnarray}
where $U$ is the internal energy and $V$ is the volume. From
classical thermodynamics, the relation between $U$, $V$ and $P$
can be written in the form \cite{Shar,Land}
\begin{eqnarray}
\frac{dU}{dV}=-P.
\end{eqnarray}
From equations (2) - (4), we get the following first order
ordinary differential equation
\begin{eqnarray}
\frac{dU}{dV}+\frac{\gamma\frac{U}{V}}{1-\beta\frac{U}{V}}=\alpha\left(\frac{U}{V}\right)^{2}.
\end{eqnarray}
Assuming the binomial expansion upto first order, we obtain the
approximate solution
\begin{eqnarray}
U\approx V\left[\frac{(\alpha+\beta)+\sqrt{(\alpha+\beta)^{2}-4 (1+\gamma)\{\alpha\beta-(\frac{V}{k})^{2(1+\gamma)}\}}}
{2\{\alpha\beta-(\frac{V}{k})^{2(1+\gamma)}\}}\right],
\end{eqnarray}
where $k$ is an integration constant ($\ne 0$) which is either
universal constant or a function of entropy ($S$). The above
solution provides the solution of energy density as
\begin{eqnarray}
\rho=\left[\frac{(\alpha+\beta)+\sqrt{(\alpha+\beta)^{2}-4
(1+\gamma)\{\alpha\beta-(\frac{V}{k})^{2(1+\gamma)}\}}}
{2\{\alpha\beta-(\frac{V}{k})^{2(1+\gamma)}\}}\right]
\end{eqnarray}
For small volumes ($V\approx 0$), the energy density of the Van
der Waals fluid behaves like
\begin{eqnarray}
\rho=\left[\frac{(\alpha+\beta)+\sqrt{(\alpha+\beta)^{2}-4
(1+\gamma)\alpha\beta}} {2\alpha\beta}\right]
\end{eqnarray}
with the condition $(\alpha+\beta)^{2} \ge 4
(1+\gamma)\alpha\beta$ and hence the minimum value of $\rho$ is
$\rho_{min}=\sqrt{\frac{1+\gamma}{\alpha\beta}}$. Now we will
discuss different physical parameters of the model.

\subsection{Pressure}
\begin{figure}
~~~~~~~~~~~~~~~~~~~~~~~~~~\includegraphics[height=1.8in]{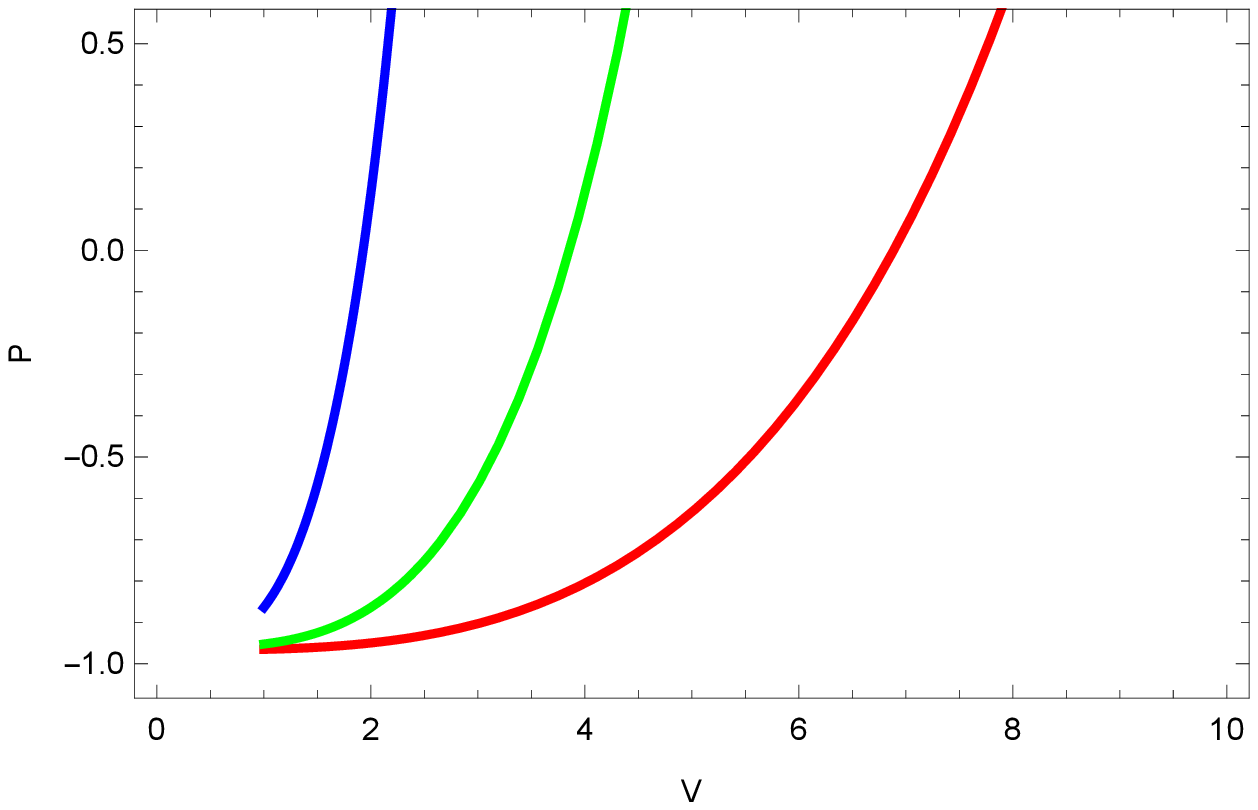}~~~~~~~~~~~~~~~~~\\

~~~~~~~~~~~~~~~~~~~~~~~~~~~~~~~~~~~~~~~~~~~~~~~~~~~~~~~~~~~~Fig.1~~~~~~~~~~~~~~~~~~~~~~~~\\

\textit{\textbf{Figure 1:} Plots of P versus V for $\gamma=0.7, \beta=1,
\alpha=20$ and $k=5$ (blue curve), $k=10$ (red curve) and $k=20$ (green curve).}\vspace{2mm}\\
\end{figure}
From equations (2) and (7), we obtain the expression pressure in
terms of $V$ as in the following form:
\begin{eqnarray}
P&=&\frac{\gamma[(\alpha+\beta)+\sqrt{(\alpha+\beta)^{2}-4
(1+\gamma)\{\alpha\beta-(\frac{V}{k})^{2(1+\gamma)}\}}]}
{\alpha\beta-[\beta^{2}+2(\frac{V}{k})^{2(1+\gamma)}
+\beta\sqrt{(\alpha+\beta)^{2}-4 (1+\gamma)\{\alpha\beta-(\frac{V}{k})^{2(1+\gamma)}\}}]}\nonumber\\
&&-\alpha\left[\frac{(\alpha+\beta)+\sqrt{(\alpha+\beta)^{2}
-4 (1+\gamma)\{\alpha\beta-(\frac{V}{k})^{2(1+\gamma)}\}}}
{2[\alpha\beta-(\frac{V}{k})^{2(1+\gamma)}]}\right]^{2}\nonumber\\
\end{eqnarray}
The trajectory of pressure given by the above equation against
volume is drawn in figure 1 for different values of $k=5,10,20$.
Figure shows the positive and negative behavior of pressure. It is
observed that the accelerating universe at small volume tends to
decelerated phase of universe at large volume.

\subsection{EoS Parameter}

From equations (2) and (7), we obtain the equation of state
parameter in terms of $V$ as in the following form:
\begin{eqnarray}
\omega=\frac{P}{\rho}&=&\frac{2\gamma[\alpha\beta-(\frac{V}{k})^{2(1+\gamma)}]}
{\alpha\beta-[\beta^{2}+2(\frac{V}{k})^{2(1+\gamma)}+\beta\sqrt{(\alpha+\beta)^{2}
-4 (1+\gamma)\{\alpha\beta-(\frac{V}{k})^{2(1+\gamma)}\}}]}\nonumber\\
&&-\alpha\left[\frac{(\alpha+\beta)+\sqrt{(\alpha+\beta)^{2}-4
(1+\gamma)\{\alpha\beta-(\frac{V}{k})^{2(1+\gamma)}\}}}
{2[\alpha\beta-(\frac{V}{k})^{2(1+\gamma)}]}\right]\nonumber\\
&=&
    \begin{cases}
      \gamma, &  V \gg k\\
      -1, & \ V\ll k .
    \end{cases}
\end{eqnarray}
\begin{figure}
~~~~~~~~~~~~~~~~~~~~~~~~~~\includegraphics[height=1.8in]{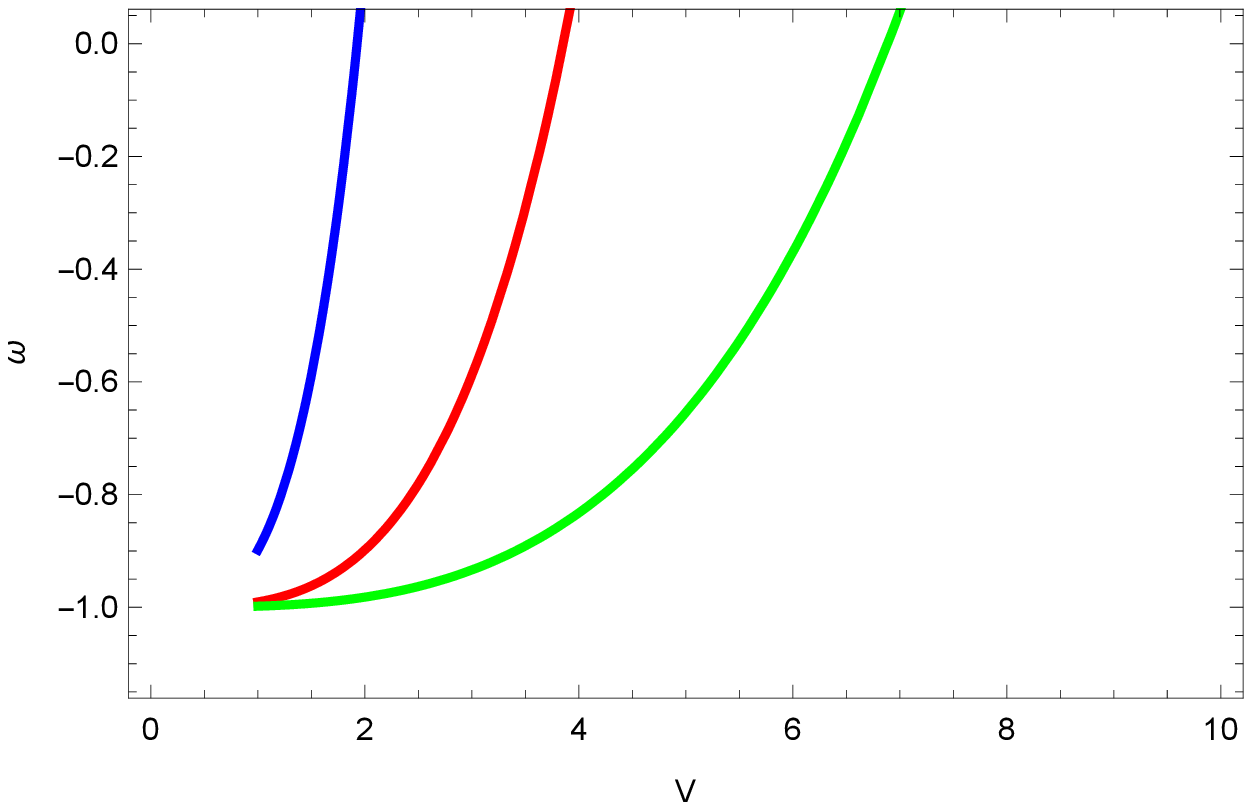}~~~~~~~~~~~~~~~~~\\

~~~~~~~~~~~~~~~~~~~~~~~~~~~~~~~~~~~~~~~~~~~~~~~~~~~~~~~~~~~~Fig.2~~~~~~~~~~~~~~~~~~~~~~~~\\

\textit{\textbf{Figure 2:} Plots of $\omega$ versus V for $\gamma=0.7, \beta=1,
\alpha=20$  $k=5$ (blue curve), $k=10$ (red curve) and $k=20$ (green curve).}\vspace{2mm}\\
\end{figure}

The EOS parameter $\omega$ is drawn in figure 2 for different
values of $k=5,10,20$. The EOS parameter transits from $-1$ to
positive values as volume increases. That means it yields
cosmological constant model for small volume, then it generates
the quintessence region and goes to positive region (tends to
$\gamma$).

\subsection{Deceleration Parameter}

The deceleration parameter is given by
\begin{eqnarray}
q=\frac{1}{2}+\frac{3P}{2\rho}&=&\frac{1}{2}+\frac{3\gamma[\alpha\beta
-(\frac{V}{k})^{2(1+\gamma)}]}{\alpha\beta-[\beta^{2}+2(\frac{V}{k})
^{2(1+\gamma)}+\beta\sqrt{(\alpha+\beta)^{2}-4 (1+\gamma)\{\alpha\beta-(\frac{V}{k})^{2(1+\gamma)}\}}]}\nonumber\\
&&-\alpha\left[\frac{(\alpha+\beta)+\sqrt{(\alpha+\beta)^{2}-4
(1+\gamma)\{\alpha\beta-(\frac{V}{k})^{2(1+\gamma)}\}}}
{2[\alpha\beta-(\frac{V}{k})^{2(1+\gamma)}]}\right]\nonumber\\
&=&
    \begin{cases}
      \frac{1}{2}+\frac{3}{2}\gamma, &  V \gg k\\
      -1, & \ V\ll k .
    \end{cases}
\end{eqnarray}
\begin{figure}

~~~~~~~~~~~~~~~~~~~~~~~~~~\includegraphics[height=1.8in]{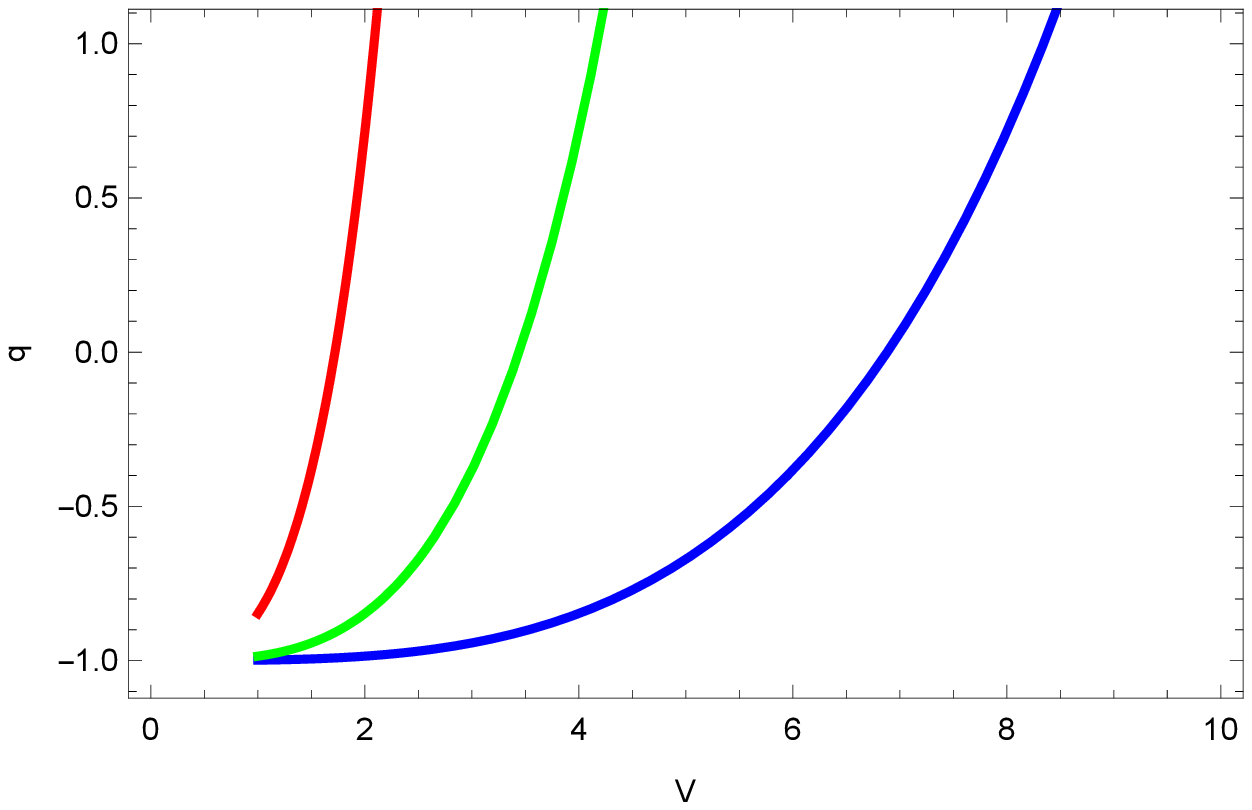}~~~~~~~~~~~~~~~~~\\

~~~~~~~~~~~~~~~~~~~~~~~~~~~~~~~~~~~~~~~~~~~~~~~~~~~~~~~~~~~~Fig.3~~~~~~~~~~~~~~~~~~~~~~~~\\

\textit{\textbf{Figure 3:}Plots of $q$ versus V for $\gamma=0.7, \beta=1,
\alpha=20$  $k=5$ (blue curve), $k=10$ (red curve) and $k=20$ (green curve).}\vspace{2mm}
\end{figure}
Figure 3 presents the trajectories of the deceleration parameter
$q$ against $V$ for different values of $k$. From the graph we can
see that $q$ increases from $-1$ to the positive value (tends to
$\frac{1+3\gamma}{2})$. It describes acceleration at small volumes
whereas at large volume, it exhibits decelerating behavior for
different values of $k$. So, a transition from accelerating to
decelerating universe is observed. It shows $q\rightarrow -1$ with
decreasing $V$ i.e it yields cosmological constant model as
$V\rightarrow 1$.

\subsection{Square Speed of Sound}
To discuss the classical stability of the model, we need to obtain
the square speed of sound which is given by
\begin{eqnarray}
V_{s}^{2}=\left(\frac{\partial P}{\partial \rho} \right)_{S}
&=&\frac{4\gamma\left[\alpha\beta-(\frac{V}{k})^{2(1+\gamma)}\right]^{2}}
{\left[\alpha\beta-\{\beta^{2}+2(\frac{V}{k})^{2(1+\gamma)}
+\beta\sqrt{(\alpha+\beta)^{2}-4 (1+\gamma)\{\alpha\beta-(\frac{V}{k})^{2(1+\gamma)}\}}\}\right]^{2}}\nonumber\\
&&-2\alpha\frac{(\alpha+\beta)+\sqrt{(\alpha+\beta)^{2}-4
(1+\gamma)\{\alpha\beta-(\frac{V}{k})^{2(1+\gamma)}\}}}
{2[\alpha\beta-(\frac{V}{k})^{2(1+\gamma)}]}\nonumber\\
&=&
    \begin{cases}
      \gamma, &  V \gg k\\\\
      4\left[\frac{\{\alpha(1+\gamma)-\beta\}\sqrt{(\alpha-\beta)^{2}-4\alpha\beta\gamma}+
      3\alpha\beta\gamma-(\alpha-\beta)^{2}}{\{\sqrt{(\alpha-\beta)^{2}-4\alpha\beta\gamma}
      -\alpha+\beta\}^{2}}\right], & \ V\ll k .
    \end{cases}
\end{eqnarray}

In figure 4 the squared speed of sound is plotted against $V$ for
different values of $k$. We observe that for $0\le V\lesssim 9$,
graph shows $V_{s}^{2}>0$ while for $V\gtrsim 9$, graph shows
$V_{s}^{2}<0$. So the model is classically stable for small volume
and for large volume it shows unstable behavior.

\begin{figure}
~~~~~~~~~~~~~~~~~~~~~~~~~~\includegraphics[height=1.8in]{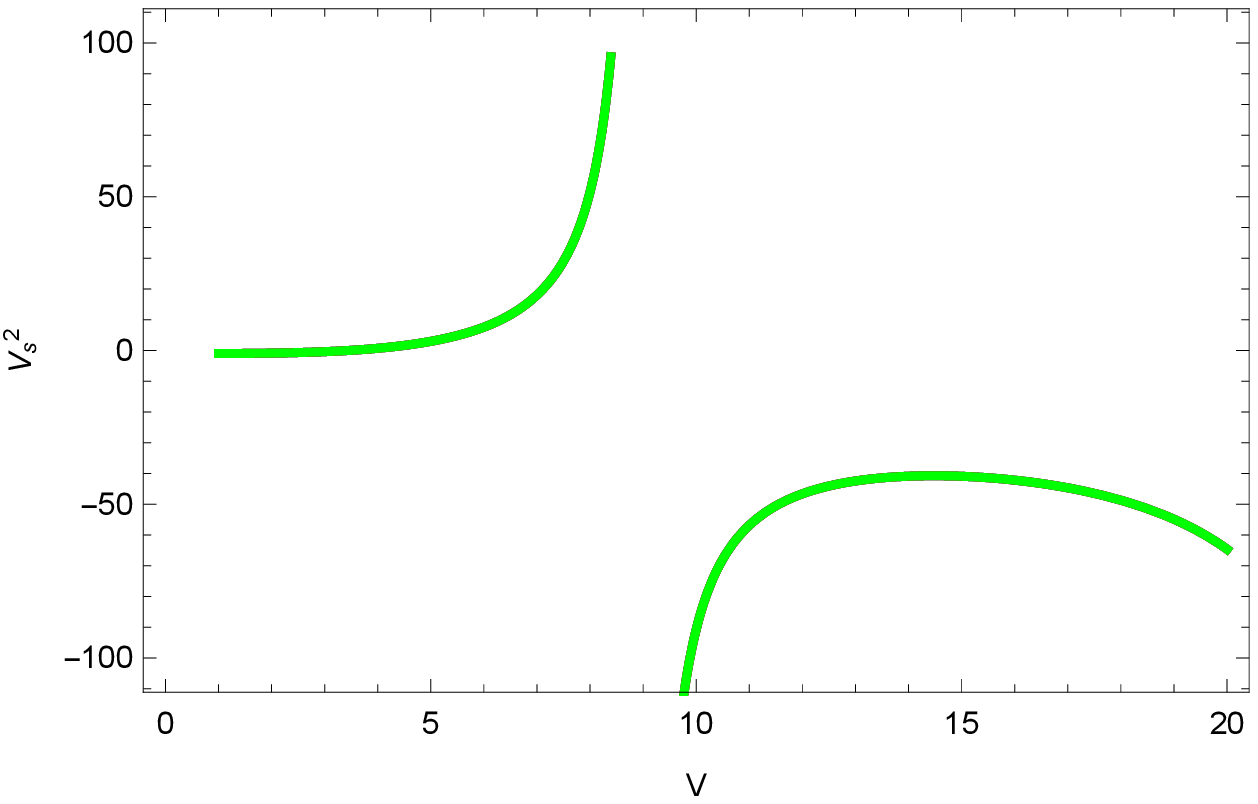}~~~~~~~~~~~~~~~~~\\

~~~~~~~~~~~~~~~~~~~~~~~~~~~~~~~~~~~~~~~~~~~~~~~~~~~~~~~~~~~~Fig.4~~~~~~~~~~~~~~~~~~~~~~~~\\

\textit{\textbf{Figure 4:}Plots of $V_{s}^{2}$ versus V for $\gamma=0.7,
\beta=1, \alpha=20 $  $k=5$ (blue curve), $k=10$ (red curve) and $k=20$ (green curve).}\vspace{2mm}\\
\end{figure}

\section{Thermodynamic Stability of the Van der Waals Fluid}

Now we will discuss the behaviour of temperature and the
thermodynamic stability of the Van der Waals Fluid. To analyze the
thermodynamic stability conditions and the evolution for the Van
der Waals fluid, it is necessary to determine (i) if the pressure
reduces through an adiabatic expansion i.e., $\left(\frac{\partial
P}{\partial V}\right)_{S}<0$, (ii) if the pressure reduces to an
expansion at constant temperature $T$ i.e., $\left(\frac{\partial
P}{\partial V}\right)_{T}<0$ and (iii) if the heat capacity at
constant volume i.e.,  $C_{V}>0$.\\

Differentiating equation (9) w.r.t. $V$, we obtain
\begin{eqnarray}
\left(\frac{\partial P}{\partial V}\right)_{S}&=&
\frac{(1+\gamma)(\frac{V}{k})^{2(1+\gamma)}}{V X^{2}\sqrt{Y}}
\left[(\alpha+\beta)\sqrt{Y}+2(1+\gamma)(\frac{V}{k})^{2(1+\gamma)}
-2\gamma\alpha\beta+\beta^{2}+\alpha^{2}\right]\nonumber\\
&&\times \left[4\gamma
X^{2}\left(\alpha\beta-(\beta^{2}+2(\frac{V}{k})^{2(1+\gamma)}+\beta\sqrt{Y})\right)^{-2}-\alpha
X^{-1}(\alpha+\beta+\sqrt{Y})\right],
\end{eqnarray}
where $X=\alpha\beta-(\frac{V}{k})^{2(1+\gamma)}$ and
$Y=(\alpha+\beta)^{2}-4(1+\gamma)X$. When the volume is very
small, the above equation reduces to zero, but for the large
volume, we have the following expression
\begin{eqnarray}
\left(\frac{\partial P}{\partial
V}\right)_{S}&&=\frac{-4(1+\gamma)\{(\beta-\alpha(1+\gamma))\sqrt{(\alpha-\beta)^{2}-4\alpha\beta\gamma}
+(\alpha-\beta)^{2}-3\alpha\beta\gamma\}}
{v \alpha^{2} \beta^{2}\{-\alpha+\beta+\sqrt{(\alpha-\beta)^{2}
-4\alpha\beta\gamma}\}\sqrt{(\alpha-\beta)^{2}-4\alpha\beta\gamma}}\nonumber\\
&&\times
\left[(\alpha+\beta)\sqrt{(\alpha-\beta)^{2}-4\alpha\beta\gamma}
+\alpha^{2}+\beta^{2}-2\alpha\beta\gamma\right].
\end{eqnarray}
Figure 5(a) shows that $\left(\frac{\partial P}{\partial
V}\right)_{S}<0$ at large volumes and at small volume, it is
positive and tends to zero as $V\rightarrow 0$. So, the adiabatic
condition is satisfied for all the considered values of $k$.\\

The specific heat capacity in constant volume is defined by
\begin{eqnarray}
C_{V}=T\left(\frac{\partial S}{\partial T}\right)_{V},
\end{eqnarray}
where the temperature can be obtained by the relation \cite{Shar}
\begin{eqnarray}
T=\frac{\partial U}{\partial S}=\left(\frac{\partial U}{\partial
k}\right) \left(\frac{\partial k}{\partial S}\right)
\end{eqnarray}
Differentiating (6) with respect to $k$, we get
\begin{eqnarray}
\frac{\partial U}{\partial k}=\frac{2(1+\gamma)(\frac{V}{k})^{2(1+\gamma)}}{k X\sqrt{Y}}[V(1+\gamma)-U \sqrt{Y}]
\end{eqnarray}
From equations (16) and (17), we obtain
\begin{eqnarray}
T=\frac{2(1+\gamma)(\frac{V}{k})^{2(1+\gamma)}}{k X\sqrt{Y}}[V(1+\gamma)
-U \sqrt{Y}]\left(\frac{\partial k}{\partial S}\right)
\end{eqnarray}
From equation (6), we can write (in the sense of dimensional
analysis) \cite{Shar}
\begin{eqnarray}
[k]^{(1+\gamma)}=[U][V]^{\gamma}
\end{eqnarray}
Using the relation $[U]=[T][S]$, we obtain
\begin{eqnarray}
[k]=[T]^{\frac{1}{(1+\gamma)}}[U]^{\frac{1}{(1+\gamma)}}[V]^{\frac{\gamma}{(1+\gamma)}}
\end{eqnarray}
From this result, we get
\begin{eqnarray}
k=(\tau \nu^{\gamma}S)^{\frac{1}{(1+\gamma)}}
\end{eqnarray}
where $\tau$ and $\nu$ are constants having the dimensions of
temperature and volume respectively. Differentiating (21), we
obtain
\begin{eqnarray}
\frac{\partial k}{\partial S}=\frac{1}{(1+\gamma)}
\left(\frac{\tau
\nu^{\gamma}}{S^{\gamma}}\right)^{\frac{1}{(1+\gamma)}}
\end{eqnarray}
Using (18) and (22) we have
\begin{eqnarray}
T=\frac{2B^{\frac{1}{1+\gamma}}S^{-\frac{\gamma}{1+\gamma}}(\frac{V}{k})
^{2(1+\gamma)}}{k X\sqrt{Y}}[V(1+\gamma)-U \sqrt{Y}]
\end{eqnarray}
where $B=\tau \nu^{\gamma}$. Using (6) and (21), the equation (23)
becomes
\begin{eqnarray}
T=-\frac{2BV^{3+2\gamma}}{X'^{2}\sqrt{Y'}}[(1+\gamma)X'+((\alpha+\beta)BS+\sqrt{Y'})\sqrt{Y'}],
\end{eqnarray}
where $X'=B^{2}S^{2}X$ and $Y'=B^{2}S^{2}Y$. When $T=0$, the
entropy $S=0$ which implies that third law of thermodynamics is
satisfied for our Van der Waals fluid model. Differentiating
eq.(24) with respect to $S$, we obtain
\begin{eqnarray}
&&\frac{\partial T}{\partial S}=\frac{2V^{3+2\gamma}B^{2}}{X'^{3}Y'^{\frac{3}{2}}}
[2BS\alpha\beta\{3X'Y'(1+\gamma)-2X'^{2}(1+\gamma)^{2}+2Y'^{2}\}+\nonumber\\
&&Y'^{\frac{3}{2}}(\alpha+\beta)\{4\alpha\beta B^{2}S^{2}-X'\}+BSX'(\alpha+\beta)^{2}\{(1+\gamma)X'+2Y'\}]
\end{eqnarray}
Now from equation (15), we have the expression of specific heat
capacity as in the following form:
\begin{eqnarray}
C_{V}=-[X'^{2}Y(1+\gamma)+X'Y'^{\frac{3}{2}}\{SB(\alpha+\beta)+\sqrt{Y}\}]
\nonumber\\
\times
\Big{(}B[2BS\alpha\beta\{3X'Y'(1+\gamma)-2X'^{2}(1+\gamma)^{2}+2Y'^{2}\}
\nonumber\\
+Y'^{\frac{3}{2}}(\alpha+\beta)\{4\alpha\beta
B^{2}S^{2}-X'\}+BSX'(\alpha+\beta)^{2}\{(1+\gamma)X'+2Y'\}]\Big{)}^{-1}
\end{eqnarray}
The specific heat $C_{V}$ is plotted as the function of volume $V$
in figure 5(b) for three different values of $k$. The positivity
of specific heat is obtained for all considered values of $k$. It
should be noted that when temperature $T$ is zero, the $C_{V}$
vanishes, which assures the validity of third law of thermodynamics.\\


\begin{figure}
\includegraphics[height=1.8in]{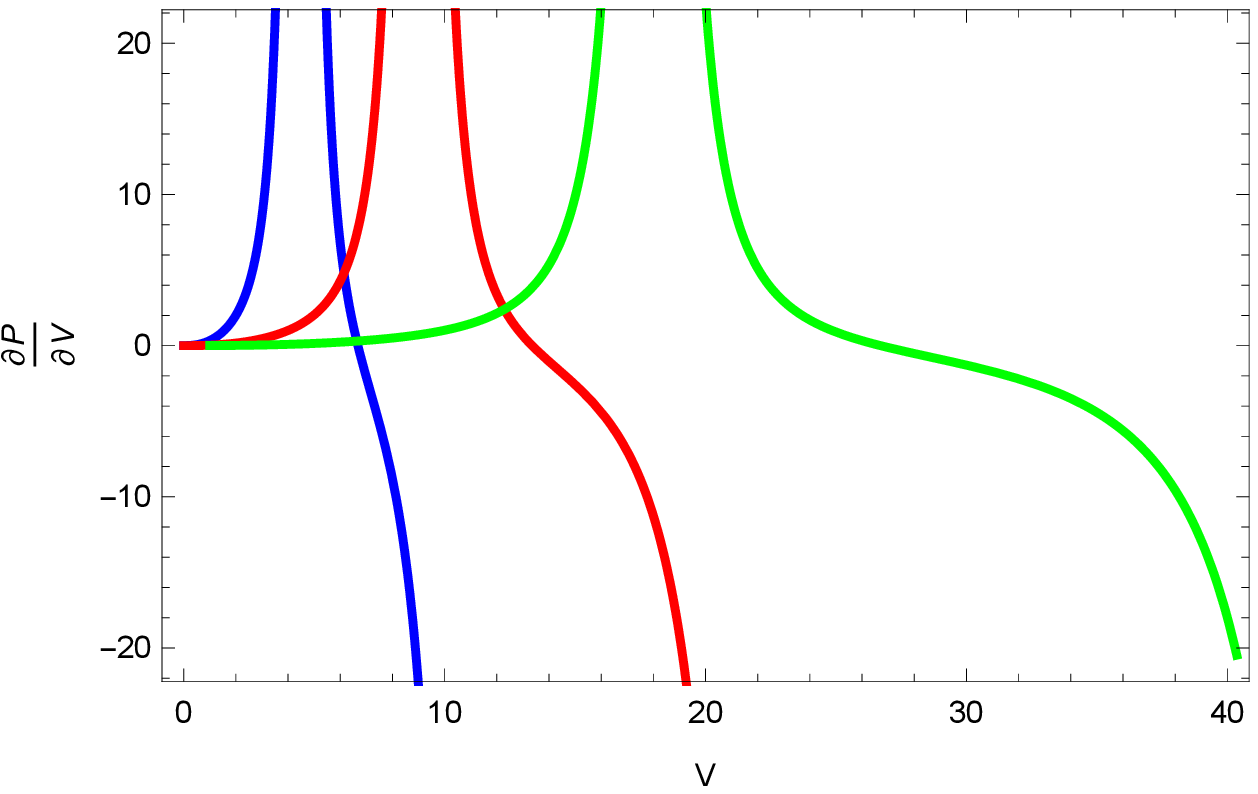}~~~\includegraphics[height=1.8in]{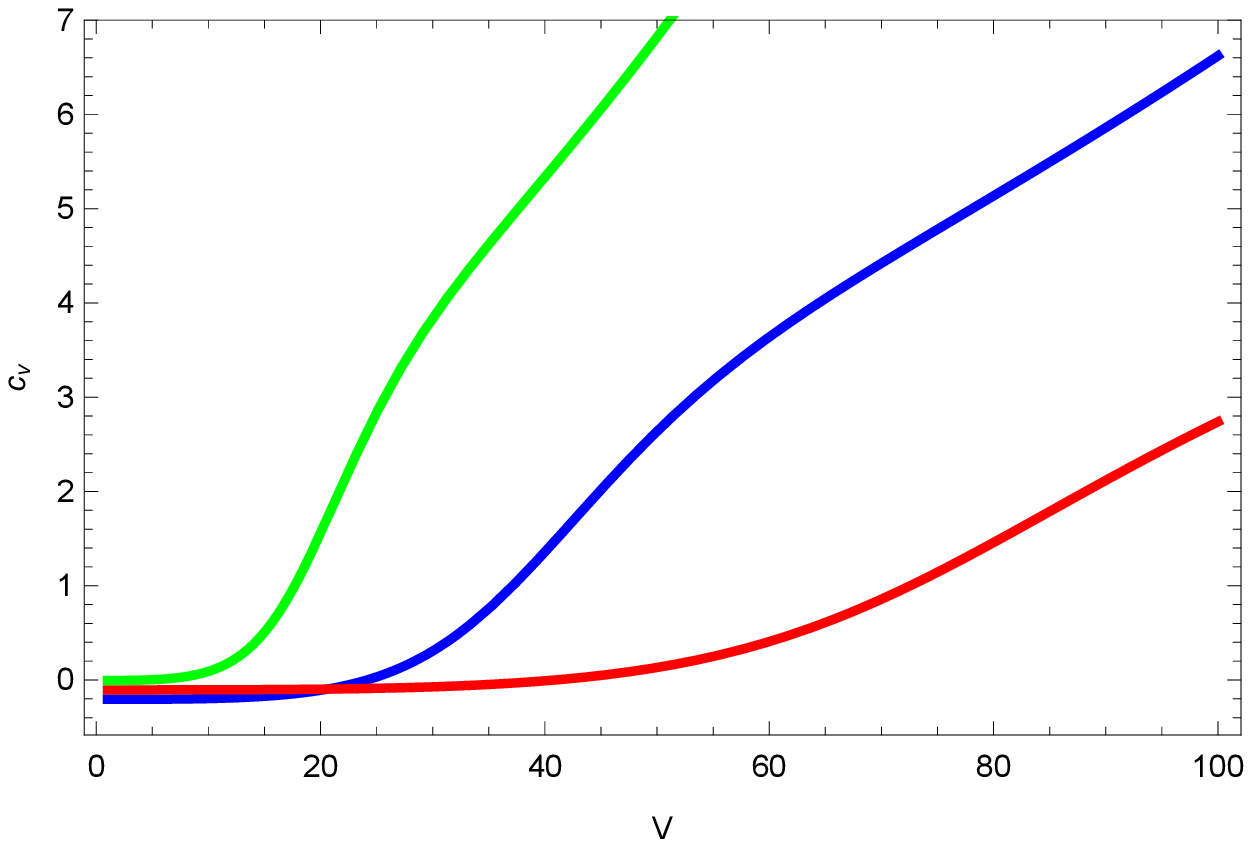}~\\

~~~~~~~~~~~~~~~~~~~~~~~~~~~~~Fig.5(a)~~~~~~~~~~~~~~~~~~~~~~~~~~~~~~~~~~~~~~~~~~~~~~~~~~~~~~Fig.5(b)~~~~~~~~~~~~~~~~\\

\textit{\textbf{Figure 5:} Plots of $\frac{\partial P}{\partial
V}$ and $C_{V}$ respectively against V for $\gamma=0.7, \beta=1,
\alpha=20 $  $k=5$ (blue curve), $k=10$ (red curve) and $k=20$
(green curve).}\vspace{2mm}
\end{figure}

\section{Discussions}

In this work, we have studied the thermodynamic properties of
cosmological fluid described by the van der Waals equation of
state in the framework of flat FRW universe. The phenomena of late
time accelerated expansion of the universe is studied through
different physical parameters like pressure, effective EoS,
deceleration parameters and squared speed of sound. Figure 1 shows
the positive and negative behavior of pressure $P$. We have also
observed that at small volume, the universe is accelerating and it
evolves to decelerated phase at large volume. From figure 2, we
have observed that the EOS parameter $\omega$ transits from $-1$
to positive values as volume increases. That means it yields
cosmological constant model for small volume, then it generates
the quintessence region and goes to positive region (tends to
$\gamma$). From figure 3, we have seen that the deceleration
parameter $q$ increases from $-1$ to the positive value (tends to
$\frac{1+3\gamma}{2})$. It describes acceleration at small volumes
whereas at large volume, it exhibits decelerating behavior. So, a
transition from accelerating to decelerating universe occurred.
That means the deceleration parameter $q$ yields cosmological
constant model for small volume and for large volume, it crosses
the quintessence region. It shows $q\rightarrow -1$ with
decreasing $V$ i.e it yields cosmological constant model as
$V\rightarrow 1$. For the stability analysis of the model, the
squared speed of sound $V_{s}^{2}$ has drawn in figure 4. We have
observed that for $0\le V\lesssim 9$, graph shows $V_{s}^{2}>0$
while for $V\gtrsim 9$, graph shows $V_{s}^{2}<0$. So the model is
classically stable for small volume and for large volume, it shows
unstable behavior. Finally, we have examined the thermodynamic
stability of the considered fluid using adiabatic, isothermal and
specific heat conditions. From figure 5(a), we have seen that
$\left(\frac{\partial P}{\partial V}\right)_{S}<0$ at large
volumes and at small volume, it is positive and tends to zero as
$V\rightarrow 0$. So, the adiabatic condition is satisfied. From
figure 5(b), we have observed that he specific heat $C_{V}$ is
always positive. So the third law of thermodynamics is obeyed for
Van der Waals fluid. For all the figures, we have assumed the
considered values of $k~(=5,10,20)$. It should be noted that when
temperature $T$ is zero, the $C_{V}$ vanishes, which assures the
validity of third law of thermodynamics.

\end{document}